\documentclass%
[aps,prb,twocolumn,
superscriptaddress,
floatfix]{revtex4}
\usepackage{amsmath}
\usepackage{hyperref}
\usepackage{amsfonts}
\usepackage{amssymb}
\usepackage[pdftex]{graphicx}
\newcommand{\fig}[2]{\includegraphics[width=#1]{#2}}

\newcommand*{\rmd}{\mathrm{d}}
\newcommand*{\dv}{\!\!\rmd v\,}

\newcommand*{\bv}[1]{\boldsymbol{#1}}

\begin{document}

\title{Minimization of phonon-tunneling dissipation in mechanical resonators}

\author{Garrett D.~Cole}\email[]{These authors contributed
  equally to this work.} 
\affiliation{Faculty of Physics, University of
  Vienna, Boltzmanngasse 5, A-1090 Vienna, Austria.}  
\author{Ignacio Wilson-Rae}\email[]{These authors contributed
  equally to this work.} 
\affiliation{Technische Universit\"{a}t M\"{u}nchen,
  85748 Garching, Germany.}
\author{Katharina Werbach}
\affiliation{Faculty of Physics, University of
  Vienna, Boltzmanngasse 5, A-1090 Vienna, Austria.}
  \author{Michael R.~Vanner} 
\affiliation{Faculty of Physics, University of
  Vienna, Boltzmanngasse 5, A-1090 Vienna, Austria.}
\author{Markus Aspelmeyer}\email[]{markus.aspelmeyer@univie.ac.at}  
\affiliation{Faculty of Physics, University of
  Vienna, Boltzmanngasse 5, A-1090 Vienna, Austria.}

\date{\today}

\begin{abstract}
  Micro- and nanoscale mechanical resonators have recently emerged as
  ubiquitous devices for use in advanced technological applications,
  for example in mobile communications and inertial sensors, and as
  novel tools for fundamental scientific endeavors. Their performance
  is in many cases limited by the deleterious effects of mechanical
  damping. Here, we report a significant advancement towards
  understanding and controlling support-induced losses in generic
  mechanical resonators. We begin by introducing an efficient
  numerical solver, based on the ``phonon-tunneling'' approach,
  capable of predicting the design-limited damping of high-quality
  mechanical resonators. Further, through careful device engineering,
  we isolate support-induced losses and perform the first rigorous
  experimental test of the strong geometric dependence of this loss
  mechanism. Our results are in excellent agreement with theory,
  demonstrating the predictive power of our approach. In combination
  with recent progress on complementary dissipation mechanisms, our
  phonon-tunneling solver represents a major step towards accurate
  prediction of the mechanical quality factor.
\end{abstract}
\pacs{\ldots}
\maketitle

Mechanical coupling of a suspended structure to its supports is a
fundamental energy loss mechanism in micro- and nanomechanical
resonators~\cite{Craighead00}. Referred to variously as
clamping~\cite{Ekinci05} or anchor loss~\cite{Wang00}, this process
remains significant even in devices fabricated from high-quality
materials operated in vacuum and at cryogenic temperatures, and is in
fact unavoidable in any non-levitating system. Although much progress
has been made towards the understanding of mechanical dissipation at
the micro- and nanoscale\cite{Cleland,Ekinci05}, obtaining reliable
predictions for the fundamental design-limited quality factor $Q$
remains a major challenge while direct experimental tests are
scarce. At the same time, the implementation of high-quality micro-
and nanomechanical systems is becoming increasingly important for
numerous advanced technological applications in sensing and metrology,
with select examples including wireless filters~\cite{Wang00,Clark05},
on-chip clocks~\cite{Lutz07},
microscopy~\cite{Sidles95,Rugar04,Degen09,Li07} and molecular-scale
mass sensing~\cite{Jensen08,Naik09}, and recently for a new generation
of macroscopic quantum experiments that involve mesoscopic mechanical
structures~\cite{Armour02,Marshall03,Blencowe04,Schwab05,Kippenberg08,Aspelmeyer08I,Aspelmeyer08II,O'Connell10}. Here
we introduce a finite-element enabled numerical solver for calculating
the support-induced losses of a broad range of low-loss mechanical
resonators. We demonstrate the efficacy of this approach via
comparison with experimental results from microfabricated devices
engineered in order to isolate support-induced losses by allowing for
a significant variation in geometry, while keeping other resonator
characteristics approximately constant. The efficiency of our solver
results from the use of a perturbative scheme that exploits the
smallness of the contact area, specifically the recently introduced
``phonon-tunneling'' approach~\cite{Wilson-Rae08}. This results in a
significant simplification over previous approaches and paves the way
for CAD-based predictive design of low-loss mechanical resonators.

The origins of mechanical damping in micro- and nanoscale systems have
been the subject of numerous studies during the last decades and
several relevant mechanisms for the decay of acoustic mechanical
excitations, i.e.~phonons, have been
investigated~\cite{Cleland,Ekinci05}. These include: (i) fundamental
anharmonic effects such as phonon-phonon interactions
\cite{Kiselev08,Cleland}, thermoelastic damping (TED)
\cite{Zener37,Lifshitz00,Duwel06,Kiselev08,Cleland}, and the Akhiezer
effect \cite{Kiselev08,Cleland}; (ii) viscous or fluidic damping,
involving interactions with the surrounding atmosphere or the
compression of thin fluidic layers
\cite{Vignola06,Karabacak07,Verbridge08}; (iii) materials losses
driven by the relaxation of intrinsic or extrinsic defects in the bulk
or surface of the resonator
\cite{Yasumura00,Mohanty02,Southworth09,Venkatesan10,Unterreithmeier10}
for which the most commonly studied model is an environment of
two-level fluctuators \cite{Seoanez08,Remus09} and (iv)
support-induced losses, i.e.~the dissipation induced by the
unavoidable coupling of the resonator to the substrate
\cite{Wang00,Mattila02,Clark05,Anetsberger08,Eichenfield09}, which
corresponds to the radiation of elastic waves into the supports
\cite{Cross01,Park04,Photiadis04,Bindel05,Wilson-Rae08,Judge07}. This
last mechanism poses a fundamental limit as vibrations of the
substrate will always be present.

These various dissipation processes add incoherently such that the
reciprocals of the corresponding $Q$-values satisfy
$1/Q_\mathrm{tot}=\sum_i 1/Q_i$, where $i$ labels the different
mechanisms. Thus, in a realistic setting care must be taken to isolate
the contribution under scrutiny. In contrast to all other damping
mechanisms (i)-(iii) which exhibit various dependencies with external
physical variables such as pressure and temperature, support-induced
dissipation is a temperature and scale-independent phenomenon with a
strong geometric character that is present in any suspended
structure. Moreover, its scale-independence implies that the same
analysis can be applied to both micro- and nanoscale devices. We
exploit this geometric character in order to isolate the
support-induced contribution and obtain a direct experimental test of
phonon-tunneling dissipation.

We first introduce our numerical solver, which provides a new
technique to efficiently model support-induced losses for a broad
class of mechanical structures. Previous approaches have relied on
either the direct solution of an elastic wave radiation problem
involving the substrate~\cite{Cross01,Park04,Photiadis04,Judge07} or
the simulation of a perfectly absorbing artificial boundary
\cite{Bindel05,Eichenfield09}, with applications typically limited to
a few specific
geometries~\cite{Bindel05,Judge07,Anetsberger08,Eichenfield09}. In
contrast, our technique represents a substantial simplification in
that it reduces the problem to the calculation of a perfectly
decoupled resonator mode together with \emph{free} elastic wave
propagation through the substrate in the absence of the suspended
structure. A key feature of our method is to combine a standard
finite-element method (FEM) calculation of the resonator mode together
with the use of an extended contact at the support. This allows us to
treat complex geometries taking proper account of interference effects
between the radiated waves.

In analogy to radiation tunneling in photonics and electron tunneling
in low-dimensional structures, we adopt a ``phonon tunneling'' picture
to describe the support-induced losses \cite{Wilson-Rae08}. In this
picture the mechanical resonance of interest, characterized by
frequency $\omega_R$, is regarded as a phonon cavity that is weakly
coupled to the exterior by a hopping process, whereby the elastic
energy leaks out of the resonator through the narrow contact areas
from which it is suspended. Within this framework, one can start from
the harmonic Hamiltonian associated with the elastic scattering
eigenmodes of the entire structure, including the substrate, and
derive a quantum model for the Brownian motion experienced by each
resonance of the suspended structure.

The corresponding weak tunnel couplings can be obtained to lowest
order in the small parameter $k_Rd$, where $1/k_R$ is the
characteristic length scale over which the resonator mode varies
appreciably and $d$ is the characteristic dimension of the contact
area $S$ from which the resonator is suspended. For typical structures
that exhibit high-$Q$ mechanical resonances, $k_Rd <<1$ is comfortably
satisfied. This justifies the weak coupling approximation and leads to
a general expression for the associated dissipation $1/Q$ in terms of
the overlaps between the scattering modes and the resonator mode. In
the limit $d\to0$ the leading contribution is obtained by replacing
the scattering modes by the free (unperturbed) modes of the supports,
which yields~\cite{Wilson-Rae08}
\begin{align}\label{Q}
  \frac{1}{Q} \,=\, & \frac{\pi}{2 \rho_s
    \rho_R\omega^3_R} \int_q \left| \int_S \rmd \bar{S} \cdot
    \left(\bv{\sigma}^{(0)}_q \cdot \bar{u}'_R - \bv{\sigma}'_R
      \cdot \bar{u}^{(0)}_q \right) \right|^2 \nonumber \\ &
  \times\delta [\omega_R - \omega(q) ]\,.
\end{align}
Here $\bv{\sigma}'_{R}$ and $\bar{u}'_{R}$ are the stress and
displacement fields associated with the normalized resonator mode,
$\bv{\sigma}^{(0)}_{q}$ and $\bar{u}^{(0)}_{q}$ are the analogous
fields for the continuum of support modes labeled by $q$
[eigenfrequencies $\omega(q)$], and $\rho_s$ and $\rho_R$ are,
respectively, the densities of the substrate and resonator
materials. The resonator mode should satisfy either (i) free or (ii)
clamped boundary conditions at the contact area $S$ depending on the
behavior of the eigenmode when $S$ is small, while the unperturbed
support modes should satisfy the converse. These homogeneous boundary
conditions correspond, respectively, to $\rmd \bar{S}
\cdot\bv{\sigma}'_R=0$ and $\bar{u}'_R=0$ so that only one of the two
terms in the surface integral is finite. Physical examples of case (i)
are pedestal geometries, such as microspheres, microdisks, or
microtoroids, when the pedestal has good impedance match with the
substrate and is thus included in the support. On the other hand
examples of case (ii) include the planar structures investigated here,
where the supports consist of the portion of the structure that is not
free-standing. Further details about the scope and validity of
Eq.~(\ref{Q}) are given in the SI Appendix~\ref{app:solver}.

\begin{figure}
\fig{0.48\textwidth}{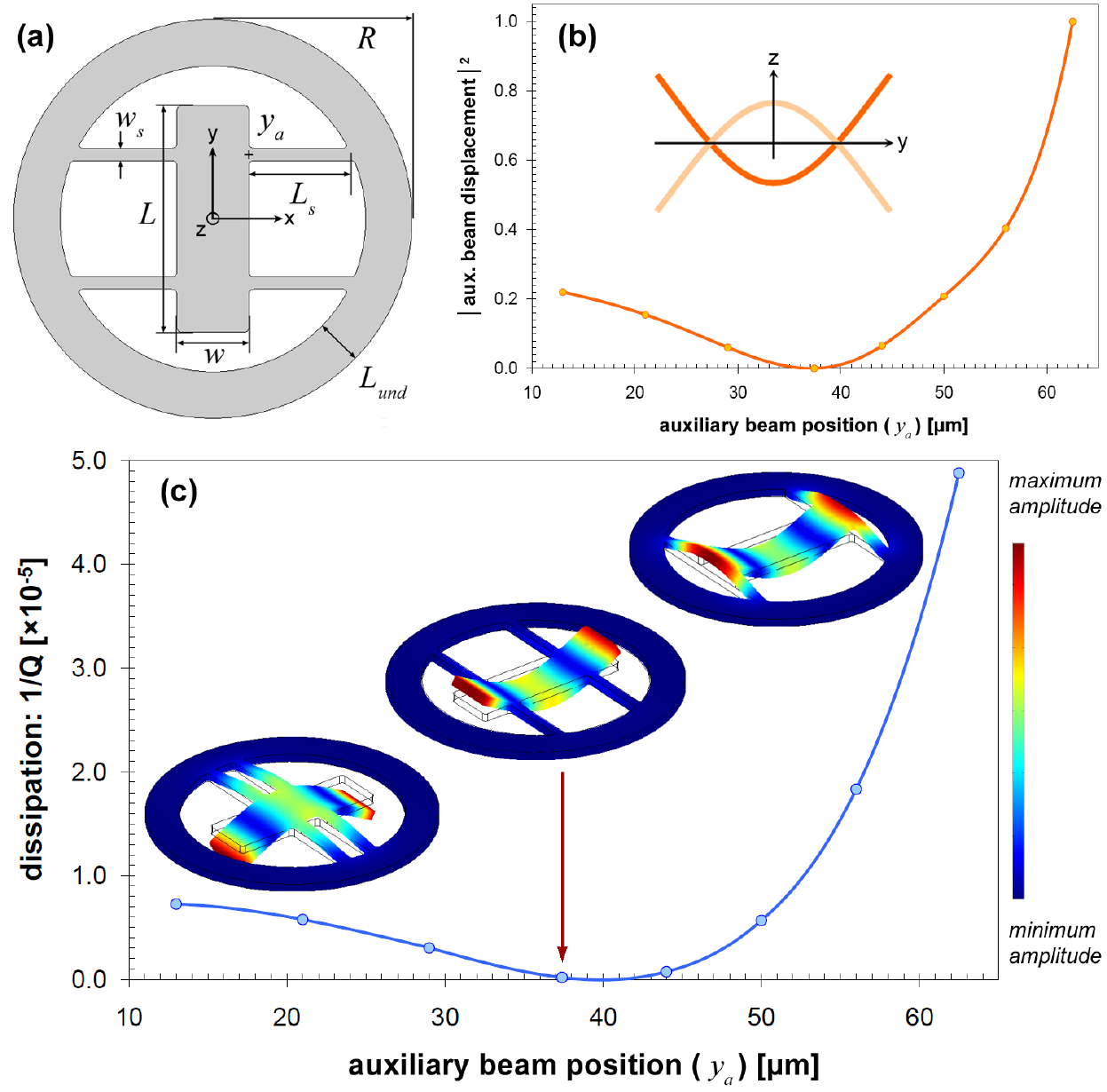}
\caption{Mapping out phonon-tunneling dissipation in a free-free
  resonator. (a) Schematic diagram of the resonator geometry. (b)
  Normalized squared center of mass displacement of a single
  auxiliary-beam-central-resonator contact calculated via FEM (the
  inset shows the profile of the free-free mode as approximated by
  Euler-Bernoulli theory). (c) Simulated dissipation
  [cf.~Eq.~\ref{Qinscribed}] as a function of the auxiliary beam's
  $y$-coordinate ($y_a$). Values corresponding to 8 discrete
  geometries were calculated here with $\mathsf{t}=6.67\,\mu$m,
  $\mathsf{w}_s=7\,\mu$m, $\mathsf{w}=42\,\mu$m, $L=132\,\mu$m,
  $R=116\,\mu$m, and $L_\textrm{und}=27\,\mu$m---the line is simply a
  guide for the eye. The FEM calculated mode shapes correspond to the
  three extreme examples of the resonator design, from left to right:
  auxiliary beams near the resonator center ($y_a=13\,\mu$m), beams
  near the ideal nodal position ($y_a=37.4\,\mu$m), and beams attached
  at the ends ($y_a=62.5\,\mu$m). The theoretical clamping loss limit
  $1/Q_\mathrm{th}$ for nodal positioning is always finite with the
  geometry closest to this position (indicated by the arrow) yielding
  $1/Q_\mathrm{th}\approx 2\times10^{-7}$. }
\end{figure}

Though the aforementioned framework is completely general, in order to
investigate the predictive power of our approach, we focus
specifically on the flexural modes of a symmetric plate geometry of
thickness $\mathsf{t}$ that is inscribed in a circle of radius $R$,
with the contact area $S$ corresponding to the outer rim of an
idealized circular undercut (undercut distance of
$L_\textrm{und}$). To calculate the theoretical $Q$-values of such
devices via Eq.~(\ref{Q}) we have developed a numerical solution
technique that determines the resonator eigenmode and eigenfrequency
via FEM (with $\bar{u}'_R=0$ at $S$) and uses a decomposition into
cylindrical modes for the support, which is approximated by the
substrate modelled as an isotropic elastic half-space. The latter
approximation is expected to be quantitatively precise for the
low-lying flexural resonances when the underetched gap between the
suspended structure and the substrate satisfies $h<R$ (where $h$ is
the gap height), and the largest resonant wavelength for elastic wave
propagation in the substrate is smaller than the relevant length
scales characterizing the mounting of the sample (see below). The
aforementioned weak coupling condition, $k_Rd <<1$, follows in this
case from $\mathsf{t}\ll R$. From Eq.~(\ref{Q}) we obtain [cf.~SI
Appendix~\ref{app:solver} for details of this derivation]
\begin{equation}\label{Qinscribed}
 \frac{1}{Q} \,=\,  \frac{\pi}{2 \rho_s
    \rho_R\omega_R}\sum_{n,\gamma}\frac{|f_{z,n}|^2}{c_\gamma^3}\,
\tilde{u}_{n,\gamma}(\omega_RR/c_\gamma,\nu_s)\,, 
\end{equation}
where we use cylindrical coordinates and introduce the dimensionless
displacements $\tilde{u}_{n,\gamma}(\tilde
q,\nu_s)=2\pi\int_0^{\pi/2}\rmd \theta \sin\theta |u^{(0)}_{\bar
  q,\gamma;z}(0,\nu_s)|^2J^2_n(\tilde q \sin\theta)$ and the linear
stress Fourier components $f_{z,n}=\int_S \rmd \bar{S} \cdot
\bv{\sigma}'_R \cdot \hat{z} e^{i n \phi}$ with
$n=0,\pm1,\pm2,\ldots$. Here $\gamma=l$, $t$, $s$ labels the different
types of relevant plane-wave modes $\bar{u}^{(0)}_{\bar
  q,\gamma;z}(\bar r,\nu_s)$ of the half-space \cite{Graff}
[i.e.~longitudinal ($l$), transverse SV ($t$), and surface acoustic
waves ($s$) given that transverse SH waves do not contribute] with
$c_\gamma$ the corresponding speed of sound --- as determined by the
density $\rho_s$, Poisson ratio $\nu_s$, and Young's modulus $E_s$ of
the substrate. We adopt spherical coordinates for the incident
wave vector $\bar q$ with polar angle $\theta$. It is straightforward
to generalize the above to in-plane modes and to plate geometries
inscribed in a rectangle. Furthermore, we stress that the rim need not
be continuous, as in cases where the resonator volume contacts the
support at a disjoint set of small areas.

\begin{figure*}
\fig{\textwidth}{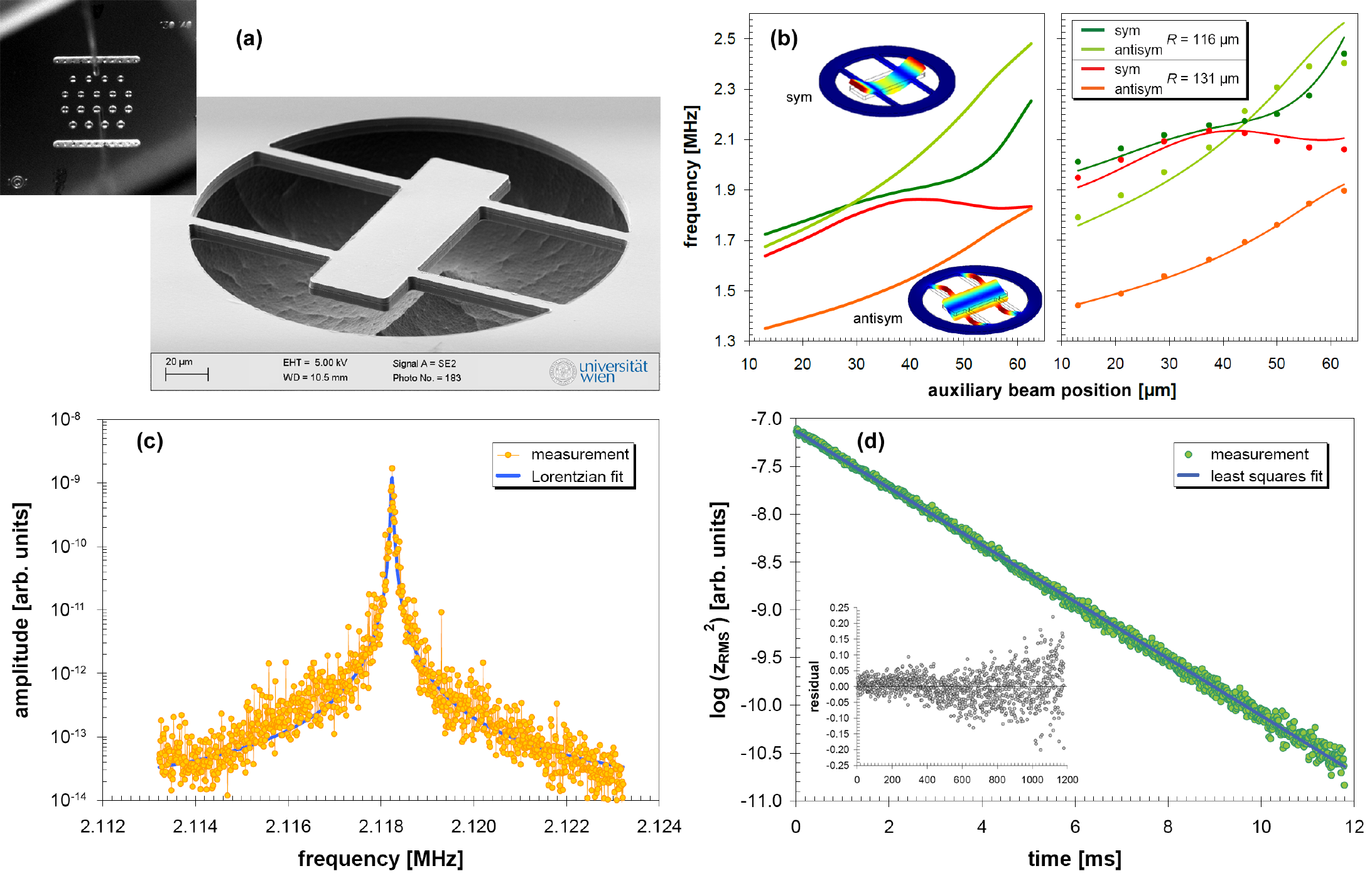}
\caption{Characterization of the completed free-free resonators. (a)
  Optical micrograph of the $5\,$mm$\times5\,$mm chip containing the
  batch-fabricated microresonators as well as an electron micrograph
  highlighting a single suspended structure. (b) Simulated (left) and
  measured (right) eigenfrequencies as a function of the auxiliary
  beam $y$-coordinate. The measured values (discrete points) show
  excellent agreement with the simulated dataset, albeit with a slight
  offset dependent on the parity of the mode. The fitting lines in the
  right plot correspond to a mean frequency offset of $262\,$kHz for
  the symmetric free-free modes and $89\,$kHz for the neighboring
  antisymmetric modes (inset shows the FEM derived mode shape of the
  antisymmetric resonance). Lower panel --- examples of the fitting
  techniques utilized for $Q$-value extraction including: (c)
  Lorentzian fitting of the free-free resonance (captured on a
  spectrum analyzer) for a device with $R=116\,\mu$m and
  $y_a=29\,\mu$m resulting in $Q=4.5\times10^4$; and (d) ringdown
  fitting of the same device using linear regression of the natural
  log of the mean-square of the free-ringdown signal captured
  single-shot with a high speed oscilloscope yielding
  $Q=4.46\times10^4$. The inset includes the residuals to the linear
  fit showing an excellent agreement with the expected exponential
  decay.}
\end{figure*}

For experimental verification of our solver, we have developed
``free-free'' micromechanical resonators consisting of a central
plate (resonator) of length $L$ and width $\mathsf{w}$ suspended by
four auxiliary beams as depicted in Fig.~1(a). These structures are
etched from a high-reflectivity monocrystalline distributed Bragg
reflector (DBR) --- as described in the Methods Section, suited for
Fabry-Perot-based optomechanical systems \cite{Groblacher09}. The
devices used in this study constitute a variant of the previously
demonstrated free-free flexural design in which auxiliary beams with
widths $\mathsf{w}_s\ll\mathsf{w}$ and lengths $L_s=\lambda_t/4$
(where $\lambda_t$ is the resonant wavelength for the propagation of
torsional waves) placed at the nodes of the central resonator mode
provide noise filters to suppress support-induced losses
\cite{Wang00}. A major drawback with the $\lambda_t/4$-beam design is
that the resulting auxiliary beam length can be excessive. In fact for
the eigenfrequencies investigated in this work, the corresponding beam
length ($>400\mu$m at $1.7$MHz) leads to a proliferation of low
frequency flexural resonances that compromise the stability of the
optical cavity and render mode identification difficult. We circumvent this
issue by utilizing instead a reduced auxiliary beam length
$L_s\ll\lambda_t/4$ chosen to avoid matching flexural resonances.

The free-free design provides an ideal platform to isolate and measure
phonon tunneling dissipation: first, by altering the attachment
position of the auxiliary beams, this design allows for a significant
variation of geometry, while approximately preserving the frequencies
and effective surface-to-volume ratios of the resonators. As these
characteristics are kept constant, one can rule out the influence of
additional damping mechanisms (specifically those driven by internal
losses and surface effects) on the variation in $Q$ and hence isolate
support-induced losses in the measured devices. Second, the free-free
resonators provide an intuitive illustration of the strong geometric
character of support-induced dissipation. Intuitively, the clamping
loss will be proportional to the elastic energy radiated through the
auxiliary beams which should approximately scale as the squared
deflection of their contacts with the central resonator [cf.~Fig.~1
panels (b) and (c)]. Thus varying the contact position of the
auxiliary beams results in a characteristic modulation of the damping
rate which approximately maps out the central resonator mode shape. As
expected, the minimum-loss design corresponds to the geometry in which
the auxiliary beams are attached at the nodes of the fundamental
resonance of the central resonator. It is interesting to note that the
theoretical clamping loss limit $1/Q_\mathrm{th}$ for nodal
positioning is always \emph{finite} as described in [Fig.~1(c)].

For mode identification we compare the optically measured resonator
frequencies (as a function of the auxiliary beam position) with the
theoretical eigenfrequency response.  The simulated values are
generated using the geometric parameters determined via careful
analysis of the completed resonators (cf.  SI Appendix
\ref{App:device-analysis}). As can be seen in Fig.~2(b), in addition
to the symmetric free-free resonance, there is also an antisymmetric
eigenmode with comparable frequency.  We observe no mode-coupling
between these resonances, which is consistent with the specific mirror
symmetries of the structure. The frequencies are accurately reproduced
by the FEM simulation if we allow for frequency offsets that are
solely dependent on the mode parity ($262\,$kHz offset for the free-free
mode and $89\,$kHz offset for the anti-symmetric mode). We attribute
these shifts to a materials-related dissipation mechanism, involving
both surface and bulk contributions (see SI Appendix \ref{App:meas}
for further details).

All dissipation measurements have been performed at high vacuum
($\approx10^{-7}$ mbar) and at cryogenic temperatures ($20$ K) in
order to suppress fluidic and thermoelastic damping in the
devices. Under these conditions, we record quality factors spanning
$1.4\times10^4$ to $5.1\times10^4$ (cf.~Fig.~3), with the minimum $Q$
corresponding to the free-free mode of devices with an auxiliary
position of $62.5\mu$m and $R =116\mu$m, and with the maximum $Q$ to
the geometry closest to nodal positioning ($37.4\mu$m) for the same
radius and type of mode. For the symmetric mode, we readily observe
the expected characteristic modulation in $Q$ as a function of the
placement of the auxiliary beams (cf.~Fig.~1) with a relative
variation of $\Delta Q_\mathrm{exp}/Q_\mathrm{exp} \sim260\%$
($\sim80\%$) for $R=116\mu$m ($R=131\mu$m). At the same time, the
frequency variation is kept small ``qua design'', with a range of
$\Delta f/f \sim20\%$ ($\sim10\%$). In contrast the $Q$-values for the
antisymmetric mode are nearly constant with $Q\approx2.1\times10^4$
[Fig.~3(c)]. This is expected as the theoretical support-induced loss
for this mode is negligible. Additionally, as this resonance involves
mainly deformations of the auxiliary beams, its dissipation is not
expected to be correlated with the mode shapes of the central
resonator. The damping of this mode is instead dominated by other
sources of dissipation, most likely by the materials-related losses
that are also responsible for the frequency shifts. Thus we obtain an
independent corroboration that the characteristic $Q$-variation
observed for the free-free mode is indeed induced by the modification
of the geometry rather than by the small frequency variation present
in the devices.

\begin{figure}
\fig{0.4\textwidth}{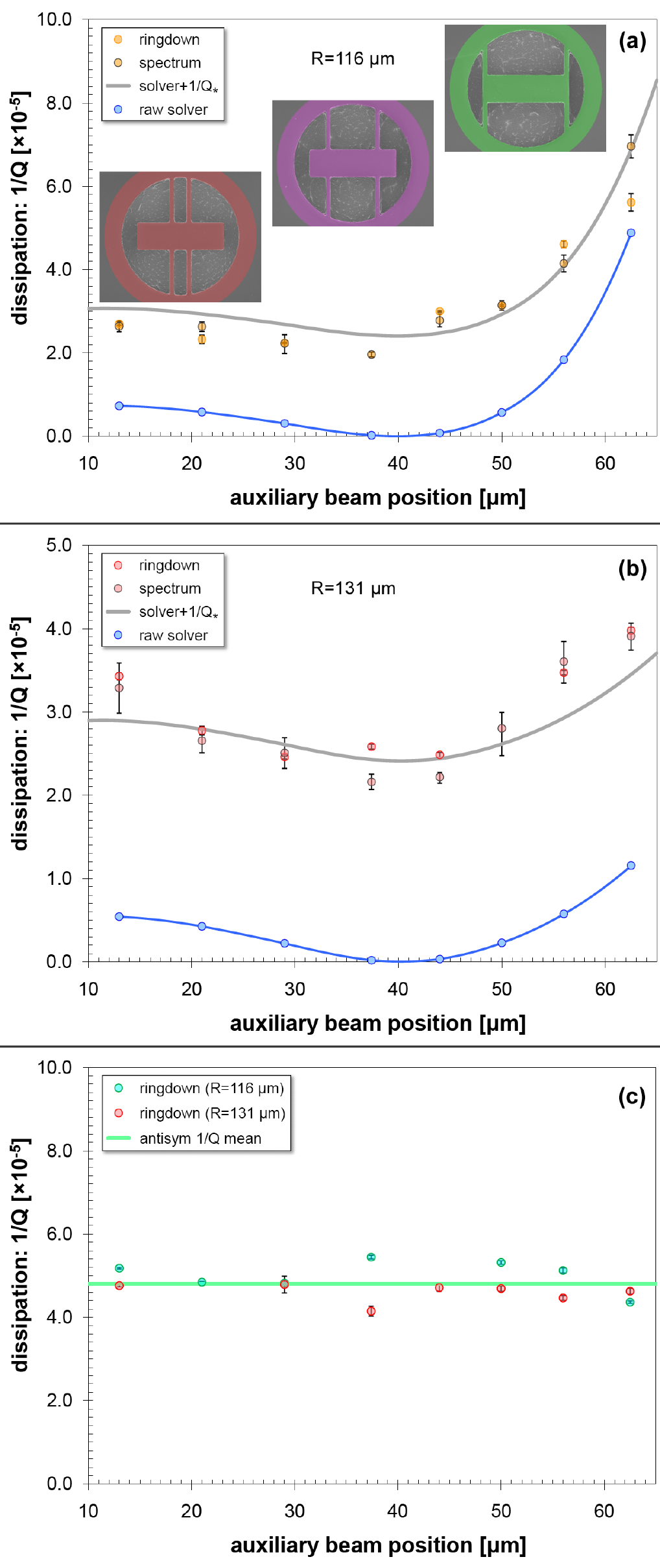}
\caption{Compiled dissipation results displaying excellent agreement
  between theory and experiment. (a) and (b) Comparison of
  experimental measurements at $T=20\,$K with theoretical dissipation
  values for the free-free mode of resonators with measured central
  dimensions of $132\,\mu$m$\times42\,\mu$m and radius $R = 116\,\mu$m
  and $R = 131\,\mu$m, respectively. Both ringdown and
  spectrally-derived data are included, with values averaged over two
  nominally-identical chips. We include both raw simulated data as
  well as fitted data (continuous lines are a guide to the eye)
  incorporating a constant offset $1/Q_*=2.41\times10^{-5}$. For the
  effective substrate we utilize the mechanical properties of Ti,
  which is the main constituent of the positioning system on which the
  chips are mounted ($\rho = 4540\,$kg/m$^3$, $E_s=116\,$GPa, and
  $\nu_s=0.34$). (c) Measured dissipation for the antisymmetric mode
  of the same structures exhibiting a lack of geometric dependence.}
\end{figure}


In order to quantitatively compare the measurements with our numerical
predictions, two additional issues must be considered: (i) our model
only captures support-induced losses, while other loss mechanisms may
still contribute to the overall damping in the devices, and (ii) the
parameters for the half-space model of the substrate must be properly
chosen. Consideration (i), together with the fact that we have
designed sets of resonators for which the frequencies and effective
surface-to-volume ratios are kept approximately constant, implies that
any additional damping mechanism that is relevant at low temperatures
and high vacuum, but is insensitive to the variation in geometry,
should contribute a constant offset $1/Q_*$ in the measured
dissipation $1/Q_\textrm{tot}$. Consideration (ii) is non-trivial
given the long-wavelength nature of the elastic waves radiated into
the substrate. For an average resonator frequency $\approx 2.12\,$MHz
estimates of the maximum wavelength for the freely-propagating elastic
waves yield a value of $\approx 2.5\,$mm, which largely exceeds the
wafer thickness ($300\pm25\,\mu$m). Thus, the mechanical materials
parameters for the substrate should be determined by the properties of
the underlying stage and positioning mechanism in the cryostat rather
than by those of the chip itself. Hence, we assume for the half-space
the mechanical properties of polycrystalline commercially-pure (grade
2) titanium (cf.~Fig.~3), of which the bulk of the structure beneath
the resonator consists. Taking all of this into account, the theory
shows remarkable agreement with the measured dissipation (as shown in
Fig.~3). It is important to note that the only free parameter used in
the model of the free-free mode is a constant offset of
$1/Q_*=2.41\times10^{-5}$. Although the exact nature of the
corresponding dissipation mechanism is currently unknown, we assume
that it arises from materials losses in the resonator
epi-structure. As our design isolates the support induced-losses,
these results establish the first systematic test of phonon tunneling
dissipation in mechanical resonators.

It should be noted that most commercially viable resonators operate in
a regime where TED dominates, and in some instances intuitive
understandings of the support-induced
damping~\cite{Wang00,Clark05,Mattila02} have allowed for its
suppression below other limiting damping mechanisms. Nonetheless, if
current efforts to minimize TED in such structures at room temperature
are successful~\cite{Duwel06}, support-induced losses may pose the
next challenge for maximizing $Q$. On the other hand, in fundamental
research thrusts employing high vacuum and cryogenic systems,
support-induced losses can become a dominant factor
\cite{Anetsberger08,Eichenfield09}. For example, the free-free designs
explored here provide a route to minimize support-induced losses for
application in optomechanical experiments utilizing the
micromechanical resonator as an end mirror in a high-finesse
Fabry-Perot cavity \cite{Groblacher09}. To gauge the relevance of our
``free-free'' micromirror design in this context, it is instructive to
compare the fundamental limit at nodal positioning
$Q_\mathrm{th}\approx 5\times10^{6}$ and the maximum $Q$-value
measured of $5.1\times10^4$, with the corresponding results for the
fundamental flexural mode of a clamped bridge of comparable
dimensions. In fact for the typical dimensions considered, as required
for integration in a high-performance Fabry-Perot cavity, we obtain a
theoretical limit $Q_{c-c}\sim10^3$ in line with prior measurements on
monocrystalline DBR optomechanical structures \cite{Cole08}.

Given the scale independent nature of support-induced losses, our
solver applies equally well to nanoscale mechanical devices. We find
that for a recent demonstration of a nanomechanical doubly-clamped
beam coupled to a superconducting qubit at mK
temperatures~\cite{LaHaye09}, the measured values for the resonator's
maximum $Q$ ($\approx6\times10^4$) can be understood solely via the
phonon-tunneling loss model (beam geometry of $0.3\mu$m$\times 0.18
\mu$m$ \times 6\mu$m) which predicts a $Q$ of $5.4\times10^4$, in
excellent agreement with the experimental value.

In conclusion, we have developed an efficient FEM-enabled numerical
method for predicting the support-induced dissipation in micro- and
nanoscale mechanical resonators. In combination with existing models
for other relevant damping channels (e.g.~fluidic and TED
\cite{Lifshitz00,Duwel06}), our ``phonon-tunneling'' solver makes
further strides towards accurate prediction of the mechanical quality
$Q$. Furthermore, we demonstrate a stringent experimental test of the
corresponding theory using resonators engineered to isolate
support-induced losses. Our results demonstrate unambiguously that
phonon-tunneling plays a significant role in the mechanical
dissipation of these devices and illustrates the strong geometric
character of this fundamental damping mechanism. Finally, we note that
the weak coupling approximation underlying our treatment is more
general than the condition of small contact area. Thus, our numerical
solver can in principle be extended to other relevant scenarios such
as systems in which the stresses at the contact are strongly
suppressed, as for example phononic-band-gap
structures~\cite{Eichenfield09}.

GDC is a recipient of a Marie Curie Fellowship of the European
Commission (EC). Additional financial support is provided by the EC
(projects MINOS, IQOS), the Austrian Science Fund (projects START,
L426, SFB FoQuS), and the European Research Council (ERC StG
QOM). Microfabrication was carried out at the Zentrum f\"{u}r Mikro-
und Nanostrukturen (ZMNS) of the Technische Universit\"{a}t Wien. GDC
gratefully acknowledges Stephan Puchegger and Markus Schinnerl for
assistance with scanning electron microscopy and focused ion beam
milling. IWR acknowledges financial support via the Nanosystems
Initiative Munich, KW via the Austrian Research Promotion Agency
(FFG).

\section{Methods}

{\bf Epitaxial Materials Structure and Resonator Fabrication
  Procedure} -- The materials structure for our high reflectivity
resonators consists of $40.5$-periods of alternating quarter-wave GaAs
(high index) and AlAs (low index) layers grown lattice-matched to an
off-cut monocrystalline germanium substrate. The ideal total thickness
of the heterostructure is $6857.6\,$nm, with individual layer
thicknesses of $77.6\,$nm and $91.9\,$nm for the GaAs and AlAs
respectively; yielding a nominal peak reflectivity at 1064 nm, as with
our previous optomechanics experiments~\cite{Cole08}. With this
design, the germanium substrate enables the use of a high-selectivity
gas-phase etching procedure, based on the noble-gas halide XeF2, in
order to rapidly and selectively undercut the underlying germanium
substrate. Thus we realize a free-standing epitaxial Bragg mirror via
a simple and fast-turnaround fabrication procedure. The details of
both the epitaxial materials design and microfabrication procedure are
covered in Ref.~\onlinecite{Cole10}.

{\bf Measurement technique} -- To characterize the frequency response
of our microresonators we utilize a custom-built optical fiber
interferometer featuring a continuous flow $^{4}$He cryostat as the sample
chamber~\cite{Cole10II}. High sensitivity displacement resolution is
achieved in this system via optical homodyne interferometry. Cryogenic
testing of these devices is necessitated due to the limitations
imposed by TED at room temperature. Estimations of the magnitude of
TED is possible using the analytical and finite element models
developed previously \cite{Zener37,Lifshitz00,Duwel06}, which predict
a $Q$-value of $\sim4000$ for the current DBR composition and thickness at
$1.8$ MHz and 300 K --- consistent with performed measurements. In
order to avoid TED our cryostat enables interrogation down to 20 K
(resulting in an estimated TED limited $Q$ of $9.9\times10^{8}$); the
minimum temperature is currently limited by the large view-port above
the sample stage. Additionally, this system is capable of vacuum
levels down to $2.5\times10^{-7}$ millibar at cryogenic temperatures,
removing any additional damping induced by fluidic, or squeeze film
effects \cite{Vignola06,Karabacak07,Verbridge08}. The eigenmodes of
the resonator are excited by driving a high-frequency (10 MHz) piezo
disc soldered to a copper stage in thermal contact with the cold
finger. For spectral characterization, the piezo disc is driven with
white noise and the resonator frequency response is recorded on a
low-noise spectrum analyzer. For the free-ringdown measurements, the
decay of a resonantly excited device is recorded in a single shot on a
high-speed oscilloscope.

\section{Supplementary information}

\appendix

\section{Numerical calculation of \texorpdfstring{$Q$}\
  -values}\label{app:solver} 

A rigorous derivation of Eq.~(1) is given in
Ref.~\onlinecite{Wilson-Rae08}. Alternatively, if one uses a
decomposition of the displacement field in terms of the unperturbed
support modes and the discrete modes of the resonator volume,
Eq.~(\ref{Q}) follows simply from applying Fermi's Golden rule to
phonon decay with the interaction Hamiltonian between the resonator
volume (labeled $<$) and the surrounding supports (labeled $>$) given
by $\int_S \rmd \bar{S}\cdot\bv{\sigma}_> \cdot \bar{u}_<$ for case
(i) and $-\int_S \rmd \bar{S} \cdot\bv{\sigma}_< \cdot \bar{u}_>$ for
case (ii). Within this framework, it is straightforward to realize
that the validity of Eq.~(\ref{Q}) is more general than the condition
$k_Rd\ll1$ and will also apply to systems in which the stresses at the
contact are exponentially suppressed, like for example
phononic-band-gap structures. In general the decomposition between
``resonator volume'' and ``supports'' consistent with the weak
coupling condition need not be unique. For our case the use of this
master formula is completely equivalent to prior intuitive approaches
based on forcing the substrate with the stress source generated by the
resonator, as can be shown rigorously by using \textemdash for the
elastic Green's function of the substrate \textemdash a spectral
decomposition in terms of its free modes. Finally, one should note
that Eq.~(\ref{Q}) assumes that the resonance is spectrally resolved
and that the support-induced frequency shifts are small compared with
the free spectral range \cite{Wilson-Rae08}. Thus, in the presence of
mode-coupling \cite{Bindel05,Anetsberger08} our treatment remains
valid provided that the mode-coupling is not dominated by
support-induced interactions, which includes the case where it is
accounted for by FEM assuming perfect clamping and excludes cases
where symmetry-breaking induced by the support is relevant. In the
particular case studied here, the suspended structure has reflection
symmetries with respect to the $x-y$, $x-z$, and $y-z$ planes; while
the entire structure including the substrate is only invariant under
the latter two, allowing one to classify the modes accordingly.

To derive Eq.~(\ref{Qinscribed}) from Eq.~(\ref{Q}) we adopt for
the free elastic half-space, modelling the decoupled support, a
decomposition into eigenmodes
$\bar{u}^{(0)}_{q,\theta,n,\gamma}(\bar{r})$ (with
$n=0,\pm1,\pm2,\ldots$ and $q>0$) that have axial symmetry with
respect to $z$ (cf.~Fig.~1). These are related to the plane wave
eigenmodes $\bar{u}^{(0)}_{\bar{q},\gamma}(\bar{r})$ by
\begin{equation}\label{cylindrical}
  \bar{u}^{(0)}_{q,\theta,n,\gamma}(\bar{r})=\frac{1}{\sqrt{2\pi}}\int_{-\pi}^{\pi}
  \rmd\varphi\,\left(-i\right)^ne^{in\varphi}
  \bar{u}^{(0)}_{\bar{q}(q,\theta,\varphi),\gamma}(\bar{r})  
\end{equation}
where $\gamma=l$, $t$, $s$ labels the different types of relevant
modes [i.e.~longitudinal ($l$), transverse SV ($t$), and SAW ($s$)
given that SH waves do not contribute] and we adopt spherical
coordinates for the incident wavevector
$\bar{q}(q,\theta,\varphi)=q(\sin\theta\cos\varphi,\sin\theta\sin\varphi,
\cos\theta)$ [$\theta=\pi/2$ for $\gamma=s$ and $\theta\leq\pi/2$
otherwise]. We note that for the suspended plate geometry considered,
the appropriate resonator mode satisfies $\bar{u}'_R=0$ at the contact
$S$ so that we need to evaluate the second term in Eq.~(\ref{Q}). 

The thin plate condition $\mathsf{t}\ll R$ directly allows us, given
the flexural nature of the modes of interest, to neglect stresses at
$S$ that are parallel to the substrate with the possible exception of
bending-moment contributions \cite{Graff} --- this also applies if
there are small transverse dimensions comparable to
$\mathsf{t}$. However the bending-moment contributions also become
negligible in the limit $\mathsf{t}/R\to0$ as can be shown by using:
(i) that given $\omega_R \ll c_\gamma/R$ $\forall\,\gamma$, we can
Taylor expand $\bar{u}^{(0)}_{q,\theta,n,\gamma}(\bar{r})$ at the
origin in the integral over $S$, (ii) that we can assume relevant
stresses to be concentrated around the ends of the auxiliary beams so
that the bending moments at $S$ are mostly oriented along $y$, (iii)
the reflection symmetries, and (iv) that, barring interference
effects, these bending-moment contributions are at most of relative
order \cite{Wilson-Rae08} $k_R \mathsf{t}$ --- here
$k_R=(12\rho_R/E_R)^{1/4}\sqrt{\omega_R/\mathsf{t}}$ is the resonant
wavevector for the propagation of flexural waves. Thus, we find that
for all mode types other than $-+$ [antisymmetric (symmetric) with
respect to $R_x$ ($R_y$)] the correction associated to neglecting the
bending moments scales as $\Delta Q/Q\sim(k_R \mathsf{t})^2$, while
for $-+$ modes it scales as $\Delta Q/Q\sim k_R \mathsf{t}$ (note that
$L\sim R$). In turn, we find that the relative error in using
Eq.~(\ref{Q}), arising from the weak coupling approximation, scales in
this case as $\Delta
Q/Q\sim|\Delta\omega_R|/\omega_R\sim|\Delta_I(\omega_R)|/\omega_R\sim(k_R
\mathsf{t})^3$, where the phonon-tunneling induced frequency shift is
approximated by
$\Delta_I(\omega_R)\approx-(2/\pi)\int_0^\infty\rmd\omega
I(\omega)/\omega$ with $I(\omega)$ the environmental spectrum
\cite{Wilson-Rae08}.

Hence we can assume $\rmd\bar{S}\cdot\bv{\sigma}'_{R}\parallel\hat z$
and neglect the variation of
$\bar{u}^{(0)}_{q,\theta,n,\gamma}(\bar{r})$ across the thickness
$\mathsf{t}$ (i.e.~the $z$-dependence at $S$), so that the support
modes only enter into Eq.~(\ref{Q}) through
$u^{(0)}_{q,\theta,n,\gamma;z}(\bar{r})|_{z=0}$. To determine the
latter we adopt cylindrical coordinates $\bar
r=(r\cos\phi,r\sin\phi,z)$, substitute into Eq.~(\ref{cylindrical})
the plane wave expressions for
$u^{(0)}_{\bar{q}(q,\theta,\varphi),\gamma;z}(\bar{r})$, exploit that
reflection at the free surface preserves the tangential component of
the wavevector, and use the Bessel integral
$J_n(x)=\frac{1}{2\pi}\int_{-\pi}^{\pi}e^{-i(n\phi-x\sin\phi)}\rmd\phi$. Thus
we obtain
\begin{widetext}
\begin{equation}
  u^{(0)}_{q,\theta,n,\gamma;z}(\bar{r})|_{z=0}=
  \frac{(-i)^n}{\sqrt{2\pi}}
 u^{(0)}_{\bar{q}(q,\theta,\varphi),\gamma;z}(0)
  \int_{-\pi}^{\pi}
 e^{-i\left[n\varphi+rq\sin\theta\cos\left(\phi-\varphi\right)\right]} 
  \rmd\varphi = \sqrt{2\pi}
  u^{(0)}_{\bar{q}(q,\theta,\varphi),\gamma;z}\!(0) J_n(rq\sin\theta)
  e^{i n\phi}\,, 
\end{equation}
that substituted into $\hat z\cdot\bar{u}^{(0)}_{q}(\bar{r})$ in
Eq.~(\ref{Q}) leads to Eq.~(\ref{Qinscribed}) after using
$\int_q\to\sum_{n,\gamma}\int_0^{\infty}\rmd q q^{d_\gamma-1}$ $
\int_0^{\pi/2}\rmd\theta (\pi/2)^{d_\gamma-3}(\sin\theta)^{d_\gamma-2}\theta$
($d_\gamma$ is the dimensionality, i.e.~$d_\gamma=2$ for $\gamma=s$
and $d_\gamma=3$ for $\gamma\neq s$), performing the substitution
$\omega=c_\gamma q$ (for each $\gamma$), and integrating over
$\omega$. Subsequently, substitution of the explicit expressions for
the plane wave eigenmodes $\bar{u}^{(0)}_{\bar{q},\gamma}(\bar{r})$
(see for example Appendix A in Ref.~\onlinecite{Wilson-Rae08}) and
$v=\cos\theta$ into the definition of $\tilde{u}_{n,\gamma}(\tilde
q,\nu_s)$ allows us to obtain:
\begin{align}
 \tilde{u}_{n,l}(\tilde q,\nu_s) =\,
  & \left. \frac{1}{\pi^2}\int_0^1\dv \frac{(1 - 2 \alpha +
      2 \alpha v^2)^2 v^2J_n^2(\tilde q\sqrt{1-v^2})}{\left[ 4
        \alpha^{3/2} \sqrt{1-\alpha+\alpha v^2} (1-v^2) v + (1 - 2
        \alpha + 2 \alpha v^2)^2
      \right]^2}\right|_{\alpha=\alpha[\nu_s]} \nonumber\\
  \tilde{u}_{n,t}(\tilde q,\nu_s) =\, &
  \frac{4}{\pi^2} \left\{\int_0^{\sqrt{1-\alpha}}\dv
    \frac{(1-\alpha-v^2) (1-v^2) v^2J_n^2(\tilde q\sqrt{1-v^2})}{16
      (1-\alpha-v^2) (1-v^2)^2 v^2 + (2 v^2-1)^4}\right.  \nonumber
\end{align}
\begin{align}
  & \qquad +
  \left.\left.\int_{\sqrt{1-\alpha}}^{1}\dv\frac{(\alpha-1+v^2)
        (1-v^2) v^2J_n^2(\tilde q\sqrt{1-v^2})}{\left[ 4
          \sqrt{\alpha-1+v^2} (1-v^2) v + (2 v^2-1)^2 \right]^2}
    \right\}\right|_{\alpha=\alpha[\nu_s]}
  \nonumber\\
 \tilde{u}_{n,s}(\tilde q,\nu_s) = &
  \left.\frac{C^2(\alpha)}{\pi}\left[\sqrt{1-\alpha
        \xi^2\!(\alpha)} -
      \frac{1-\xi^2\!(\alpha)/2}{\sqrt{1-\xi^2\!(\alpha)}}
    \right]^{2}J_n^2(\tilde q)\right|_{\alpha=\alpha[\nu_s]} 
\end{align}
\end{widetext}
where we use the ratio $\alpha\equiv
(c_t/c_l)^2=(1-2\nu_s)/2(1-\nu_s)$ for the supports' material ($\nu_s$
is the corresponding Poisson ratio). In turn $\xi(\alpha)$ is the
ratio of the propagation velocity of surface waves to $c_t$, which is
a function of $\alpha$ that is always less than unity \cite{Graff},
and
\begin{align}
C(\alpha) & = \left\{
  \frac{2-\xi^2\!(\alpha)}{\left[1-\xi^2\!(\alpha)\right]^{3/2}}
  \left[\xi^4\!(\alpha)/4+\xi^2\!(\alpha)-1\right] \right. \nonumber
  \\ &\quad \left. + \frac{2-\alpha\xi^2\!(\alpha)}{\sqrt{1-\alpha
  \xi^2\!(\alpha)}} \right\}^{-1/2}\,.
\end{align}

The sum in Eq.~(\ref{Qinscribed}) can be reduced to a sum over
$n\geq0$ by noting that $J_{-n}(x)=(-1)J_{n}(x)$ and that as the
resonator mode is real, the linear stress Fourier components satisfy
$f_{z,-n}=f_{z,n}^*$. Furthermore, the length of the central resonator
$L$ is comparable to the radius $R$ and we focus on low-lying
resonances of the suspended structure so that the aforementioned
condition $\omega_R \ll c_\gamma/R$ $\forall\,\gamma$ is always
satisfied. This implies
$|\sum_\gamma\tilde{u}_{m,\gamma}/\sum_\gamma\tilde{u}_{n,\gamma}|\ll1$
for $m>n$ and $\sum_\gamma\tilde{u}_{n,\gamma}\neq0$, which can be
understood by considering the behavior of the Bessel functions for
small arguments.  Thus we find that in Eq.~(\ref{Qinscribed}) the sum
over the index $n$ is dominated by the first non-vanishing term as
determined by the reflection symmetries $R_x$, $R_y$. The latter also
imply ($n=0,1,2,\ldots$):
\begin{widetext}
\begin{align}
  f_{z,2n}^{\scriptscriptstyle{++}} =\, & R\int_{-\pi}^{\pi}\!\rmd\phi
  \int_{-\frac{\mathsf{t}}{2}}^{\frac{\mathsf{t}}{2}}\rmd
  z\,\hat{r}\cdot
  \bv{\sigma}^{\scriptscriptstyle{++}}_R(\bar{r})\cdot\hat{z}\cos2n\phi\,,
  \nonumber\\
  f_{z,2n+1}^{\scriptscriptstyle{+-}} =\, &
  R\int_{-\pi}^{\pi}\!\rmd\phi
  \int_{-\frac{\mathsf{t}}{2}}^{\frac{\mathsf{t}}{2}}\rmd
  z\,\hat{r}\cdot
  \bv{\sigma}^{\scriptscriptstyle{+-}}_R(\bar{r})\cdot\hat{z}\sin(2n+1)\phi\,,
  \nonumber\\
  f_{z,2n+1}^{\scriptscriptstyle{-+}} =\, &
  R\int_{-\pi}^{\pi}\!\rmd\phi
  \int_{-\frac{\mathsf{t}}{2}}^{\frac{\mathsf{t}}{2}}\rmd
  z\,\hat{r}\cdot
  \bv{\sigma}^{\scriptscriptstyle{-+}}_R(\bar{r})\cdot\hat{z}\cos(2n+1)\phi\,,
  \nonumber\\
  f_{z,2n}^{\scriptscriptstyle{--}} = & R\int_{-\pi}^{\pi}\!\rmd\phi
  \int_{-\frac{\mathsf{t}}{2}}^{\frac{\mathsf{t}}{2}}\rmd
  z\,\hat{r}\cdot
  \bv{\sigma}^{\scriptscriptstyle{--}}_R(\bar{r})\cdot\hat{z}\sin2n\phi\,,
  \nonumber\\
  f_{z,2n+1}^{\scriptscriptstyle{++}} = &
  f_{z,2n}^{\scriptscriptstyle{+-}} =\,
  f_{z,2n}^{\scriptscriptstyle{-+}} =
  f_{z,2n+1}^{\scriptscriptstyle{--}}=0 \,;
\end{align}
where the resonator mode of type $\alpha,\beta$ satisfies
$R_x\bv{\sigma}^{\alpha,\beta}_R=\alpha\bv{\sigma}^{\alpha,\beta}_R$
and
$R_y\bv{\sigma}^{\alpha,\beta}_R=\beta\bv{\sigma}^{\alpha,\beta}_R$.
To efficiently extract the above from the FEM simulation we convert
them into volume integrals using an adequate Gaussian weight so that
for example, for a fully symmetric mode we have
\begin{equation}
  f_{z,2n}^{\scriptscriptstyle{++}}=\lim_{a_*\to0}
  \frac{2}{\sqrt{\pi}a_*}\int_V\rmd r^3 
  e^{-\left(\frac{r-R}{a_*}\right)^2}
  \hat{r}\cdot\bv{\sigma}'_R(\bar{r})\cdot\hat{z}\cos2n\phi \,,
\end{equation}
\end{widetext}
where we again use cylindrical coordinates and $V$ denotes the
resonator volume. In addition we exploit that the reflection
symmetries naturally allow to perform the FEM simulation on a single
quadrant.  Thus, numerical evaluation can be conveniently performed
using a fixed $a_*$ and a mesh size $M$ such that $(V/4M)^{1/3}<
a_*\ll\mathsf{t}$. We have checked the convergence and estimate the
numerical error to be of order $5\%$.

Numerical simulations of the resonator mode are performed with the aid
of COMSOL multiphysics. Accurate three-dimensional CAD models
representing the resonator geometry are generated using Solidworks
(matched with high quality SEM images as described in Appendix
\ref{App:device-analysis}) and the bidirectional interface between the
two programs is exploited in order to perform a parametric sweep of
the auxiliary beam contact position for determining the pertinent
information about the relevant mode; namely, its eigenfrequency,
linear stress Fourier components $f_{z,n}$, and normalization
constant. In this instance a single CAD file is used with a global
variable incorporated in order to control the lateral position of the
auxiliary beams with respect to the center of the central
resonator. We use for the mechanical properties of our single-crystal
resonators an anisotropic materials model incorporating the elastic
stiffness matrix for the epitaxial structure as obtained from a
weighted average between the relative content of GaAs and AlAs
($46.37\%$ GaAs / $53.63\%$ AlAs). The corresponding parameters are:
$C_{11}=119.6\,$GPa, $C_{12}= 55.5\,$GPa, $C_{44}=59.1\,$GPa, and
$\rho_R=4483\,$kg/m$^3$. The resonator axes are aligned along $\langle
100\rangle$ (zinc-blende structure). Note that we ignore the 6-degree
misorientation of the germanium substrate as we have checked that it
has a negligible impact (error of $0.3\%$) on the simulated frequency
response of the resonators.

We note that our generic symmetric inscribed structure includes the
particular case of bridge geometries with no undercut for which a
simple variant of the method used in Ref.~\onlinecite{Wilson-Rae08}
allows us to obtain an analytical approximation for the $Q$-value of
the fundamental mode $Q_{c-c}$ valid for $3\pi\mathsf{t}/2L\ll1$. Our
scenario differs from the one considered in
Ref.~\onlinecite{Wilson-Rae08} in two ways: (i) there is now a single
half-space support instead of two and (ii) its free surface is
oriented parallel to the beam's axis instead of perpendicular to
it. Thus, for the fundamental flexural mode as the resonant wavelength
in the support is much larger than the bridge length $L$, to lowest
order in $(\mathsf{t}/L)^2$ the stresses at both clamping points add
coherently so that the overall effect of (i) is to double the
dissipation. In turn (ii) implies that the roles of the dimensionless
displacements for compression and bending are interchanged so that the
dissipation of the vertical bending modes are further corrected by a
factor of $\tilde{u}_c/\tilde{u}_v$. Thus if we consider that the
support and resonator are made of the same material characterized by a
Poisson ratio $\nu=1/3$ we obtain
\begin{equation}\label{Qcc}
 Q_{c-c}=\frac{0.92\,L^5}{\pi^4\mathsf{w}\mathsf{t}^4}\,.
\end{equation} 
Hence as a non-trivial check we have applied our numerical method to
the fundamental mode of clamped-clamped square beam monolithic geometries
with no undercut, $\nu=1/3$, and aspect ratios $L/\mathsf{t}$ ranging
from 15 to 40 and compared the results with those corresponding to
Eq.~(\ref{Qcc}).  We find a discrepancy $\epsilon$ that decreases
monotonously from $20\%$ to $4\%$ which is consistent with the rough
heuristic estimate $\epsilon\sim3\pi\mathsf{t}/2L$.

\section{Analysis of Completed Devices}\label{App:device-analysis}

The resonator layout we have designed features 16 devices on chip,
each with identical central resonator dimensions (nominally
$130\,\times\,40\,\mu$m$^2$). The 16 devices are divided into two
sub-units featuring different outer radii (116 um and 131 um
respectively), which are included in order to probe the effects of the
auxiliary beam length on the dissipation. Finally, each of the two
subsets contains 8 variations of the auxiliary beam contact position,
varying from the center to the extreme outer edge of the central beam,
with a single design chosen to match the theoretically calculated node
position (auxiliary beam positions of 13, 21, 29, $37.4$, 44, 50, 56,
and finally, $62.5\,\mu$m). To ensure a thorough investigation of each
geometry, two separate but nominally identical chips are measured.

Central to this study is an accurate determination of the geometric
properties of the optomechanical resonators. Thus, we employ a variety
of analytical techniques for the characterization of these devices as
detailed below. We find that the actual thickness of the DBR is
$6.67\,\mu$m, the central resonator dimensions are enlarged by
$1\,\mu$m at each free edge as compared with the nominal design
values, and finally, $L_\textrm{und}$ is destructively measured post
characterization and found to be on average $27\,\mu$m. In turn the
microfabrication procedure detailed Ref.~\onlinecite{Cole10} entails
$h\sim L_\textrm{und}$.

\underline{Thickness:} In order to accurately determine the physical
thickness of the resonators, we rely on measurements of the DBR
reflectance spectrum. This procedure begins by recording the
reflectance of the mirror stack (on wafer) as a function of wavelength
via spectrophotometry. A transmission matrix model is then used to fit
the measured high-reflectivity stop-band; the individual layer
thicknesses are adjusted assuming constant (fixed percentage) growth
errors for the constituent films. Note that the wavelength of peak
reflectivity of the mirror is highly sensitive to variations in layer
thickness. In fact for this structure, a 1 nm variation in the
individual layer thickness shifts the wavelength of peak reflectivity
by approximately 10 nm. Relying on accurate knowledge of the room
temperature refractive index of the binary films, we realize a minimum
wavelength resolution of $\pm1$nm; thus, the thickness accuracy is
better than 20 nm for the DBR. From this analysis we have determined
that the actual thickness of the DBR is slightly shorter than desired
at $6.67\mu$m with a peak on-wafer reflectance near 1060 nm at room
temperature (ideal target thickness of $6.86\mu$m, corresponding to a
peak wavelength of 1078 nm at 300 K).  The thickness is further
verified by scanning probe measurement of the DBR following the
anisotropic etch of the epitaxial layers. The profilometer provides an
upper limit to the DBR thickness, as additional etching arising from
surface sputtering of the Ge substrate is unavoidable. These
measurements yield a conservative thickness estimate between $6.7\mu$m
and $6.8\mu$m, verifying the more accurate spectrophotometer derived
value.

\underline{Resonator Dimensions:} The lateral dimensions of the
resonators are determined by obtaining high resolution micrographs of
each individual structure in a field emission scanning electron
microscope (Zeiss Gemini). Image analysis shows that the lateral
dimensions of the resonators have expanded by $+1\,\mu$m on each edge,
with the following results: reducing the nominal external support
diameter by $2\,\mu$m, increasing the auxiliary beam width from
$5\,\mu$m to $7\,\mu$m, and increasing the overall lateral dimensions
of the resonator by $2\,\mu$m to $132\,\mu$m and
$42\,\mu$m. Additionally, a combination of process non-idealities
(non-optimized exposure or development times) during lithography
result in the formation of a 3-$\mu$m-radius fillet of at each corner
of the device. These results are fed back into the CAD model of the
resonator in order to generate the true resonator geometry for
simulation. An overlay of the simulated resonator geometry and
micrographs obtained via scanning electron microscopy can be seen in
Fig.~3(a). Note that the resonators used in this study were not
subject to potentially damaging energetic processes beyond the
required plasma etching, including both SEM and FIB (as described
below), until all dissipation measurements had been completed.

\underline{Undercut:} In order to perform measurements of the support
undercut distance, a dual beam SEM/FIB (Zeiss Gemini) is utilized to
mill a window through the DBR and expose the underlying
germanium. Because the GaAs/AlAs heterostructure is opaque to visible
light, it is not possible to simply view the undercut distance with an
optical microscope. This method allows for an accurate determination
of the lateral etch distance below the supports. Image analysis yields
an average distance of 27 $\mu$m for the structures. Note that
multiple chips of identical geometry were released simultaneously in a
single process run, in order to ensure repeatability in the resonator
dimensions. Selected measurements across the chip verify that the
undercut length is constant for the resonators studied here (measured
values fall between $26.5$ and $28.2\,\mu$m).

\section{$Q$-value and frequency measurements}\label{App:meas}

We utilize two options for driving and characterizing the
resonance of interest: (i) by applying broadband white noise to
the piezo disc for extraction of the mechanical frequency
spectrum (simultaneously driving all modes within the system
bandwidth), and (ii) by exciting a desired mode resonantly with a
sinusoidal voltage input, abruptly shutting off the drive, and
then recording the free-ringdown of the structure. In the first
method, $Q$ is extracted by measuring the width of the
resonance of interest, while in the latter, the single-shot
amplitude decay time of the ringing structure provides the
damping rate of the resonator.

\begin{figure}
\fig{0.49\textwidth}{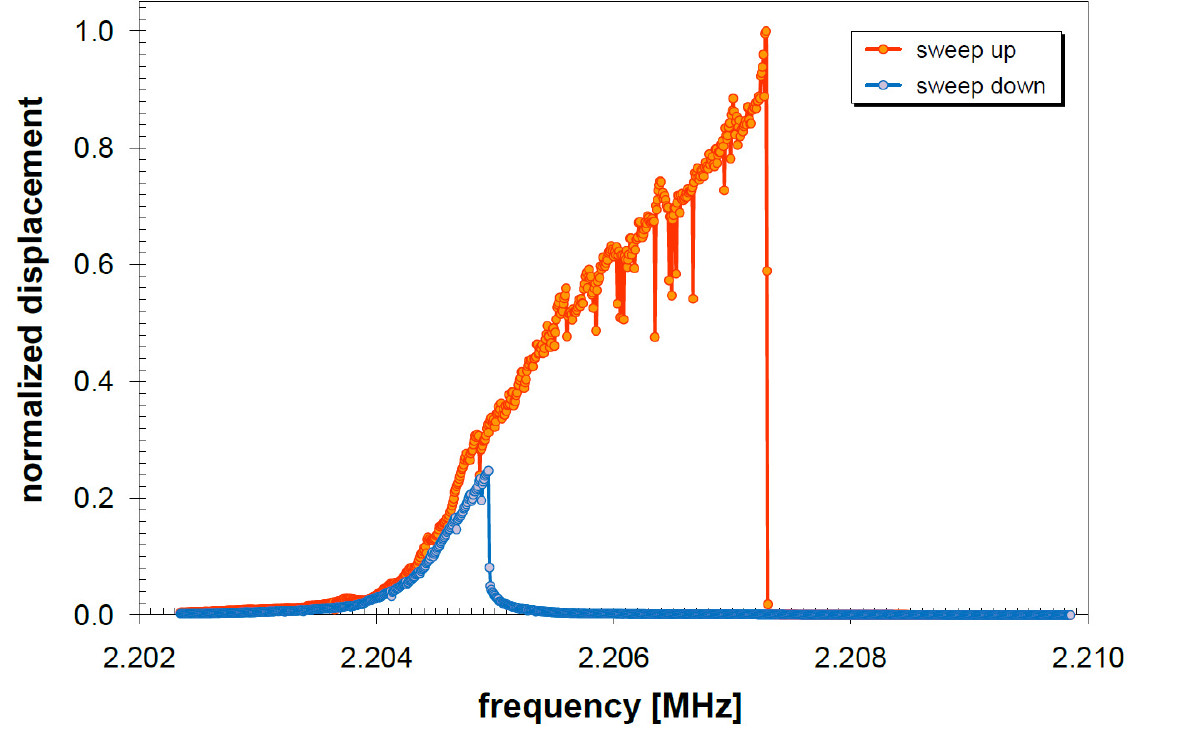}
\caption{Measured bistable response of the anti-symmetric
  mode. \label{fig:SI_nonlin}}
\end{figure}

For data analysis, we employ a combination of spectral fitting
with a Lorentzian function, with linear-regression-based fitting
of a decaying exponential in the case of the ringdown data. The
envelope of the raw ringdown signal is created by first squaring
the dataset (in order to utilize both the positive and negative
components of the decaying sinusoid) and then averaging over a
10-20 period window (it is important to note that a typical
ringdown dataset contains more than $1.5\times10^3$ periods of
oscillation). Finally, to linearize the data, we simply take the
natural logarithm of the mean-squared amplitude. In contrast to
the single-shot ringdown datasets, the spectral measurements
require multiple averages for a clean signal (typically $\sim
50-100$). The Lorentzian fit parameters include the amplitude,
center frequency, and full width at half maximum (FWHM), with
the latter two used for calculating $Q$ from their ratio. The
fast Fourier transform (FFT) of the ringdown signal can also be
used to determine the eigenfrequency, while the $1/e$ decay time
$\tau$ allows for the extraction of the resonator quality factor
via the relation $\tau=Q/\pi f$. 

Mode identification is realized by comparing the resonator frequency
response as a function of geometry with the simulated eigenfrequency
values. The modes are further distinguished by the relative
geometric-induced nonlinearity at resonance. The desired free-free
mode remains linear to the limits of our piezoelectric-based inertial
drive. On the other hand, the neighboring anti-symmetric mode exhibits
a significant hardening spring Duffing response and can readily be
driven into a bistable regime as shown in
Fig.~\ref{fig:SI_nonlin}. Care is taken to drive this mode below the
threshold for bistability to avoid complications in dissipation
extraction. This marked difference in the responses of these two types
of modes is consistent with their free-free versus clamped-clamped
nature \cite{Cleland}.

To each dissipation mechanism there is an associated dispersive effect
induced by the interactions with the corresponding environment that
shifts the resonant frequencies. For a given resonance ($\omega_R$)
this shift can be positive or negative depending on whether the
environmental spectrum is dominated, respectively, by modes with
frequencies lower or higher than $\omega_R$. In turn, the two types of
modes have markedly different surface-to-volume ratios, larger for the
antisymmetric resonance and smaller for the symmetric one; exhibit a
positive shift which is substantially larger for the symmetric mode,
and a ``background'' dissipation that is larger for the antisymmetric
one (cf.~Fig.~3). These facts can be reconciled by assuming that two
materials-related dissipation mechanisms contribute to the
``background'': a bulk one leading to an overall positive shift which
is the same for both types of modes, and a surface one leading to a
smaller negative shift that naturally scales as the surface-to-volume
ratio. In turn, theoretical estimates for the phonon-tunneling induced
shift yield a negligible \emph{negative} shift that should also follow
the mode profile leading to a significant modulation that is not
observed. Likewise mode coupling between the different resonators
would also be incompatible with a constant shift.


\begin{thebibliography}{10}
\expandafter\ifx\csname url\endcsname\relax
  \def\url#1{\texttt{#1}}\fi
\expandafter\ifx\csname urlprefix\endcsname\relax\def\urlprefix{URL }\fi
\providecommand{\bibinfo}[2]{#2}
\providecommand{\eprint}[2][]{\url{#2}}

\bibitem{Craighead00}
\bibinfo{author}{Craighead, H.~G.}
\newblock \bibinfo{title}{Nanoelectromechanical systems}.
\newblock \emph{\bibinfo{journal}{Science}} \textbf{\bibinfo{volume}{290}},
  \bibinfo{pages}{1532} (\bibinfo{year}{2000}).

\bibitem{Ekinci05}
\bibinfo{author}{Ekinci, K.~L.} \& \bibinfo{author}{Roukes, M.~L.}
\newblock \bibinfo{title}{Nanoelectromechanical systems}.
\newblock \emph{\bibinfo{journal}{Review of Scientific Instruments}}
  \textbf{\bibinfo{volume}{76}}, \bibinfo{pages}{061101}
  (\bibinfo{year}{2005}).

\bibitem{Wang00}
\bibinfo{author}{Wang, K.}, \bibinfo{author}{Wong, A.-C.} \&
  \bibinfo{author}{Nguyen, C. T.-C.}
\newblock \bibinfo{title}{VHF free-free beam high-Q micromechanical
  resonators}.
\newblock \emph{\bibinfo{journal}{J. Microelectromech. Syst.}}
  \textbf{\bibinfo{volume}{9}}, \bibinfo{pages}{347} (\bibinfo{year}{2000}).

\bibitem{Cleland}
\bibinfo{author}{Cleland, A.~N.}
\newblock \emph{\bibinfo{title}{Foundations of Nanomechanics}}
  (\bibinfo{publisher}{Springer, Berlin}, \bibinfo{year}{2003}).

\bibitem{Clark05}
\bibinfo{author}{Clark, J.}, \bibinfo{author}{Hsu, W.-T.},
  \bibinfo{author}{Abdelmoneum, M.} \& \bibinfo{author}{Nguyen, C.-C.}
\newblock \bibinfo{title}{High-Q UHF micromechanical radial-contour mode disk
  resonators}.
\newblock \emph{\bibinfo{journal}{J. Microelectromech. Syst.}}
  \textbf{\bibinfo{volume}{14}}, \bibinfo{pages}{1298--1310}
  (\bibinfo{year}{2005}).

\bibitem{Lutz07}
\bibinfo{author}{Lutz, M.} \emph{et~al.}
\newblock \bibinfo{title}{Mems oscillators for high volume commercial
  applications}.
\newblock In \emph{\bibinfo{booktitle}{Proc. Transducers, Solid-State Sensors,
  Actuators and Microsyst.}} (\bibinfo{year}{2007}).
\newblock \bibinfo{note}{14th Int. Conf.}

\bibitem{Sidles95}
\bibinfo{author}{Sidles, J.~A.} \emph{et~al.}
\newblock \bibinfo{title}{Magnetic resonance force microscopy}.
\newblock \emph{\bibinfo{journal}{Rev. Mod. Phys.}}
  \textbf{\bibinfo{volume}{67}}, \bibinfo{pages}{249} (\bibinfo{year}{1995}).

\bibitem{Rugar04}
\bibinfo{author}{Rugar, D.}, \bibinfo{author}{Budakian, R.},
  \bibinfo{author}{Mamin, H.~J.} \& \bibinfo{author}{Chui, B.~W.}
\newblock \bibinfo{title}{Single spin detection by magnetic resonance force
  microscopy}.
\newblock \emph{\bibinfo{journal}{Nature}} \textbf{\bibinfo{volume}{430}},
  \bibinfo{pages}{329--332} (\bibinfo{year}{2004}).

\bibitem{Degen09}
\bibinfo{author}{Degen, C.~L.}, \bibinfo{author}{Poggio, M.},
  \bibinfo{author}{Mamin, H.~J.}, \bibinfo{author}{Rettner, C.~T.} \&
  \bibinfo{author}{Rugar, D.}
\newblock \bibinfo{title}{Nanoscale magnetic resonance imaging}.
\newblock \emph{\bibinfo{journal}{Proc. Natl Acad. Sci. USA}}
  \textbf{\bibinfo{volume}{106}}, \bibinfo{pages}{1313--1317}
  (\bibinfo{year}{2009}).

\bibitem{Li07}
\bibinfo{author}{Li, M.}, \bibinfo{author}{Tang, H.~X.} \&
  \bibinfo{author}{Roukes, M.~L.}
\newblock \bibinfo{title}{Ultra-sensitive nems-based cantilevers for sensing,
  scanned probe and very high-frequency applications}.
\newblock \emph{\bibinfo{journal}{Nature Nanotech.}}
  \textbf{\bibinfo{volume}{2}}, \bibinfo{pages}{114--120}
  (\bibinfo{year}{2007}).

\bibitem{Jensen08}
\bibinfo{author}{Jensen, K.}, \bibinfo{author}{Kim, K.} \&
  \bibinfo{author}{Zettl, A.}
\newblock \bibinfo{title}{An atomic-resolution nanomechanical mass sensor}.
\newblock \emph{\bibinfo{journal}{Nature Nanotech.}}
  \textbf{\bibinfo{volume}{3}}, \bibinfo{pages}{533--537}
  (\bibinfo{year}{2008}).

\bibitem{Naik09}
\bibinfo{author}{Naik, A.~K.}, \bibinfo{author}{Hanay, M.~S.},
  \bibinfo{author}{Hiebert, W.~K.}, \bibinfo{author}{Feng, X.~L.} \&
  \bibinfo{author}{Roukes, M.~L.}
\newblock \bibinfo{title}{Towards single-molecule nanomechanical mass
  spectrometry}.
\newblock \emph{\bibinfo{journal}{Nature Nanotech.}}
  \textbf{\bibinfo{volume}{4}}, \bibinfo{pages}{445--450}
  (\bibinfo{year}{2009}).

\bibitem{Armour02}
\bibinfo{author}{Armour, A.~D.}, \bibinfo{author}{Blencowe, M.~P.} \&
  \bibinfo{author}{Schwab, K.~C.}
\newblock \bibinfo{title}{Entanglement and decoherence of a micromechanical
  resonator via coupling to a cooper-pair box}.
\newblock \emph{\bibinfo{journal}{Phys. Rev. Lett.}}
  \textbf{\bibinfo{volume}{88}}, \bibinfo{pages}{148301}
  (\bibinfo{year}{2002}).

\bibitem{Marshall03}
\bibinfo{author}{Marshall, W.}, \bibinfo{author}{Simon, C.},
  \bibinfo{author}{Penrose, R.} \& \bibinfo{author}{Bouwmeester, D.}
\newblock \bibinfo{title}{Towards quantum superpositions of a mirror}.
\newblock \emph{\bibinfo{journal}{Phys. Rev. Lett.}}
  \textbf{\bibinfo{volume}{91}}, \bibinfo{pages}{130401}
  (\bibinfo{year}{2003}).

\bibitem{Blencowe04}
\bibinfo{author}{Blencowe, M.}
\newblock \bibinfo{title}{Quantum electromechanical systems}.
\newblock \emph{\bibinfo{journal}{Phys. Rep.}} \textbf{\bibinfo{volume}{395}},
  \bibinfo{pages}{159} (\bibinfo{year}{2004}).

\bibitem{Schwab05}
\bibinfo{author}{Schwab, K.~C.} \& \bibinfo{author}{Roukes, M.~L.}
\newblock \bibinfo{title}{Putting mechanics into quantum mechanics}.
\newblock \emph{\bibinfo{journal}{Physics Today}}
  \textbf{\bibinfo{volume}{58}}, \bibinfo{pages}{36} (\bibinfo{year}{2005}).

\bibitem{Kippenberg08}
\bibinfo{author}{Kippenberg, T.~J.} \& \bibinfo{author}{Vahala, K.~J.}
\newblock \bibinfo{title}{Cavity optomechanics: Back-action at the mesoscale}.
\newblock \emph{\bibinfo{journal}{Science}} \textbf{\bibinfo{volume}{321}},
  \bibinfo{pages}{1172} (\bibinfo{year}{2008}).

\bibitem{Aspelmeyer08I}
\bibinfo{author}{Aspelmeyer, M.} \& \bibinfo{author}{Zeilinger, A.}
\newblock \bibinfo{title}{A quantum renaissance}.
\newblock \emph{\bibinfo{journal}{Physics World}} \textbf{\bibinfo{volume}{21}}
  (\bibinfo{year}{2008}).

\bibitem{Aspelmeyer08II}
\bibinfo{author}{Aspelmeyer, M.} \& \bibinfo{author}{Schwab, K.}
\newblock \bibinfo{title}{Focus on mechanical systems at the quantum limit}.
\newblock \emph{\bibinfo{journal}{New J. Phys.}} \textbf{\bibinfo{volume}{10}},
  \bibinfo{pages}{095001} (\bibinfo{year}{2008}).

\bibitem{O'Connell10}
\bibinfo{author}{O'Connell, A.~D.} \emph{et~al.}
\newblock \bibinfo{title}{Quantum ground state and single-phonon control of a
  mechanical resonator}.
\newblock \emph{\bibinfo{journal}{Nature}} \textbf{\bibinfo{volume}{464}},
  \bibinfo{pages}{697--703} (\bibinfo{year}{2010}).

\bibitem{Wilson-Rae08}
\bibinfo{author}{Wilson-Rae, I.}
\newblock \bibinfo{title}{Intrinsic dissipation in nanomechanical resonators
  due to phonon tunneling}.
\newblock \emph{\bibinfo{journal}{Phys. Rev. B}} \textbf{\bibinfo{volume}{77}},
  \bibinfo{pages}{245418} (\bibinfo{year}{2008}).

\bibitem{Kiselev08}
\bibinfo{author}{Kiselev, A.~A.} \& \bibinfo{author}{Iafrate, G.~J.}
\newblock \bibinfo{title}{Phonon dynamics and phonon assisted losses in
  euler-bernoulli nanobeams}.
\newblock \emph{\bibinfo{journal}{Phys. Rev. B}} \textbf{\bibinfo{volume}{77}},
  \bibinfo{pages}{205436} (\bibinfo{year}{2008}).

\bibitem{Zener37}
\bibinfo{author}{Zener, C.}
\newblock \bibinfo{title}{Internal friction in solids. i. theory of internal
  friction in reeds}.
\newblock \emph{\bibinfo{journal}{Phys. Rev.}} \textbf{\bibinfo{volume}{52}},
  \bibinfo{pages}{230} (\bibinfo{year}{1937}).

\bibitem{Lifshitz00}
\bibinfo{author}{Lifshitz, R.} \& \bibinfo{author}{Roukes, M.~L.}
\newblock \bibinfo{title}{Thermoelastic damping in micro- and nanomechanical
  systems}.
\newblock \emph{\bibinfo{journal}{Phys. Rev. B}} \textbf{\bibinfo{volume}{61}},
  \bibinfo{pages}{5600} (\bibinfo{year}{2000}).

\bibitem{Duwel06}
\bibinfo{author}{Duwel, A.}, \bibinfo{author}{Candler, R.~N.},
  \bibinfo{author}{Kenny, T.~W.} \& \bibinfo{author}{Varghese, M.}
\newblock \bibinfo{title}{Engineering mems resonators with low thermoelastic
  damping}.
\newblock \emph{\bibinfo{journal}{J. Microelectromech. Syst.}}
  \textbf{\bibinfo{volume}{15}}, \bibinfo{pages}{1437--1445}
  (\bibinfo{year}{2006}).

\bibitem{Vignola06}
\bibinfo{author}{Vignola, J.~F.} \emph{et~al.}
\newblock \bibinfo{title}{Effect of viscous loss on mechanical resonators
  designed for mass detection}.
\newblock \emph{\bibinfo{journal}{Appl. Phys. Lett.}}
  \textbf{\bibinfo{volume}{88}}, \bibinfo{pages}{041921}
  (\bibinfo{year}{2006}).

\bibitem{Karabacak07}
\bibinfo{author}{Karabacak, D.~M.}, \bibinfo{author}{Yakhot, V.} \&
  \bibinfo{author}{Ekinci, K.~L.}
\newblock \bibinfo{title}{High-frequency nanofluidics: An experimental study
  using nanomechanical resonators}.
\newblock \emph{\bibinfo{journal}{Phys. Rev. Lett.}}
  \textbf{\bibinfo{volume}{98}}, \bibinfo{pages}{254505}
  (\bibinfo{year}{2007}).

\bibitem{Verbridge08}
\bibinfo{author}{Verbridge, S.~S.}, \bibinfo{author}{Craighead, H.~G.} \&
  \bibinfo{author}{Parpia, J.~M.}
\newblock \bibinfo{title}{A megahertz nanomechanical resonator with room
  temperature quality factor over a million}.
\newblock \emph{\bibinfo{journal}{Appl. Phys. Lett.}}
  \textbf{\bibinfo{volume}{92}}, \bibinfo{pages}{013112}
  (\bibinfo{year}{2008}).

\bibitem{Yasumura00}
\bibinfo{author}{Yasumura, K.} \emph{et~al.}
\newblock \bibinfo{title}{Quality factors in micron- and submicron-thick
  cantilevers}.
\newblock \emph{\bibinfo{journal}{J. Microelectromech. Syst.}}
  \textbf{\bibinfo{volume}{9}}, \bibinfo{pages}{117 --125}
  (\bibinfo{year}{2000}).

\bibitem{Mohanty02}
\bibinfo{author}{Mohanty, P.} \emph{et~al.}
\newblock \bibinfo{title}{Intrinsic dissipation in high-frequency
  micromechanical resonators}.
\newblock \emph{\bibinfo{journal}{Phys. Rev. B}} \textbf{\bibinfo{volume}{66}},
  \bibinfo{pages}{085416} (\bibinfo{year}{2002}).

\bibitem{Southworth09}
\bibinfo{author}{Southworth, D.~R.} \emph{et~al.}
\newblock \bibinfo{title}{Stress and silicon nitride: A crack in the universal
  dissipation of glasses}.
\newblock \emph{\bibinfo{journal}{Phys. Rev. Lett.}}
  \textbf{\bibinfo{volume}{102}}, \bibinfo{pages}{225503}
  (\bibinfo{year}{2009}).

\bibitem{Venkatesan10}
\bibinfo{author}{Venkatesan, A.} \emph{et~al.}
\newblock \bibinfo{title}{Dissipation due to tunneling two-level systems in
  gold nanomechanical resonators}.
\newblock \emph{\bibinfo{journal}{Phys. Rev. B}} \textbf{\bibinfo{volume}{81}},
  \bibinfo{pages}{073410} (\bibinfo{year}{2010}).

\bibitem{Unterreithmeier10}
\bibinfo{author}{Unterreithmeier, Q.~P.}, \bibinfo{author}{Faust, T.} \&
  \bibinfo{author}{Kotthaus, J.~P.}
\newblock \bibinfo{title}{Damping of nanomechanical resonators}.
\newblock \emph{\bibinfo{journal}{Phys. Rev. Lett.}}
  \textbf{\bibinfo{volume}{105}}, \bibinfo{pages}{027205}
  (\bibinfo{year}{2010}).

\bibitem{Seoanez08}
\bibinfo{author}{Seo\'anez, C.}, \bibinfo{author}{Guinea, F.} \&
  \bibinfo{author}{Castro~Neto, A.~H.}
\newblock \bibinfo{title}{Surface dissipation in nanoelectromechanical systems:
  Unified description with the standard tunneling model and effects of metallic
  electrodes}.
\newblock \emph{\bibinfo{journal}{Phys. Rev. B}} \textbf{\bibinfo{volume}{77}},
  \bibinfo{pages}{125107} (\bibinfo{year}{2008}).

\bibitem{Remus09}
\bibinfo{author}{Remus, L.~G.}, \bibinfo{author}{Blencowe, M.~P.} \&
  \bibinfo{author}{Tanaka, Y.}
\newblock \bibinfo{title}{Damping and decoherence of a nanomechanical resonator
  due to a few two-level systems}.
\newblock \emph{\bibinfo{journal}{Phys. Rev. B}} \textbf{\bibinfo{volume}{80}},
  \bibinfo{pages}{174103} (\bibinfo{year}{2009}).

\bibitem{Mattila02}
\bibinfo{author}{Mattila, T.} \emph{et~al.}
\newblock \bibinfo{title}{A 12 MHz micromechanical bulk acoustic mode
  oscillator}.
\newblock \emph{\bibinfo{journal}{Sensors and Actuators A: Physical}}
  \textbf{\bibinfo{volume}{101}}, \bibinfo{pages}{1--9} (\bibinfo{year}{2002}).

\bibitem{Anetsberger08}
\bibinfo{author}{Anetsberger, G.}, \bibinfo{author}{Riviere, R.},
  \bibinfo{author}{Schliesser, A.}, \bibinfo{author}{Arcizet, O.} \&
  \bibinfo{author}{Kippenberg, T.~J.}
\newblock \bibinfo{title}{Ultralow-dissipation optomechanical resonators on a
  chip}.
\newblock \emph{\bibinfo{journal}{Nature Photon.}}
  \textbf{\bibinfo{volume}{2}}, \bibinfo{pages}{627--633}
  (\bibinfo{year}{2008}).

\bibitem{Eichenfield09}
\bibinfo{author}{Eichenfield, M.}, \bibinfo{author}{Chan, J.},
  \bibinfo{author}{Camacho, R.~M.}, \bibinfo{author}{Vahala, K.~J.} \&
  \bibinfo{author}{Painter, O.}
\newblock \bibinfo{title}{Optomechanical crystals}.
\newblock \emph{\bibinfo{journal}{Nature}} \textbf{\bibinfo{volume}{462}},
  \bibinfo{pages}{78--82} (\bibinfo{year}{2009}).

\bibitem{Cross01}
\bibinfo{author}{Cross, M.~C.} \& \bibinfo{author}{Lifshitz, R.}
\newblock \bibinfo{title}{Elastic wave transmission at an abrupt junction in a
  thin plate with application to heat transport and vibrations in mesoscopic
  systems}.
\newblock \emph{\bibinfo{journal}{Phys. Rev. B}} \textbf{\bibinfo{volume}{64}},
  \bibinfo{pages}{085324} (\bibinfo{year}{2001}).

\bibitem{Park04}
\bibinfo{author}{Park, Y.-H.} \& \bibinfo{author}{Park, K.~C.}
\newblock \bibinfo{title}{High-fidelity modeling of MEMS
  resonators -- Part I:
  Anchor loss mechanisms through substrate}.
\newblock \emph{\bibinfo{journal}{J. Microelectromech. Syst.}}
  \textbf{\bibinfo{volume}{13}}, \bibinfo{pages}{238} (\bibinfo{year}{2004}).

\bibitem{Photiadis04}
\bibinfo{author}{Photiadis, D.~M.} \& \bibinfo{author}{Judge, J.~A.}
\newblock \bibinfo{title}{Attachment losses of high Q oscillators}.
\newblock \emph{\bibinfo{journal}{Appl. Phys. Lett.}}
  \textbf{\bibinfo{volume}{85}}, \bibinfo{pages}{482} (\bibinfo{year}{2004}).

\bibitem{Bindel05}
\bibinfo{author}{Bindel, D.~S.} \& \bibinfo{author}{Govindjee, S.}
\newblock \bibinfo{title}{Elastic PMLs for resonator anchor loss simulation}.
\newblock \emph{\bibinfo{journal}{Int. J. Numer. Meth. Engng.}}
  \textbf{\bibinfo{volume}{64}}, \bibinfo{pages}{789} (\bibinfo{year}{2005}).

\bibitem{Judge07}
\bibinfo{author}{Judge, J.~A.}, \bibinfo{author}{Photiadis, D.~M.},
  \bibinfo{author}{Vignola, J.~F.}, \bibinfo{author}{Houston, B.~H.} \&
  \bibinfo{author}{Jarzynski, J.}
\newblock \bibinfo{title}{Attachment loss of micromechanical and nanomechanical
  resonators in the limits of thick and thin support structures}.
\newblock \emph{\bibinfo{journal}{J. Appl. Phys.}}
  \textbf{\bibinfo{volume}{101}}, \bibinfo{pages}{013521}
  (\bibinfo{year}{2007}).

\bibitem{Graff}
\bibinfo{author}{Graff, K.~F.}
\newblock \emph{\bibinfo{title}{Wave Motion in Elastic Solids}}
  (\bibinfo{publisher}{Dover, New York}, \bibinfo{year}{1991}).

\bibitem{Groblacher09}
\bibinfo{author}{Groblacher, S.} \emph{et~al.}
\newblock \bibinfo{title}{Demonstration of an ultracold micro-optomechanical
  oscillator in a cryogenic cavity}.
\newblock \emph{\bibinfo{journal}{Nature Phys.}} \textbf{\bibinfo{volume}{5}},
  \bibinfo{pages}{485--488} (\bibinfo{year}{2009}).

\bibitem{Cole08}
\bibinfo{author}{Cole, G.~D.}, \bibinfo{author}{Groblacher, S.},
  \bibinfo{author}{Gugler, K.}, \bibinfo{author}{Gigan, S.} \&
  \bibinfo{author}{Aspelmeyer, M.}
\newblock \bibinfo{title}{Monocrystalline Al(x)Ga(1-x)As heterostructures for
  high-reflectivity high-Q micromechanical resonators in the megahertz regime}.
\newblock \emph{\bibinfo{journal}{Appl. Phys. Lett.}}
  \textbf{\bibinfo{volume}{92}}, \bibinfo{pages}{261108}
  (\bibinfo{year}{2008}).

\bibitem{LaHaye09}
\bibinfo{author}{LaHaye, M.~D.}, \bibinfo{author}{Suh, J.},
  \bibinfo{author}{Echternach, P.~M.}, \bibinfo{author}{Schwab, K.~C.} \&
  \bibinfo{author}{Roukes, M.~L.}
\newblock \bibinfo{title}{Nanomechanical measurements of a superconducting
  qubit}.
\newblock \emph{\bibinfo{journal}{Nature}} \textbf{\bibinfo{volume}{459}},
  \bibinfo{pages}{960--964} (\bibinfo{year}{2009}).

\bibitem{Cole10}
\bibinfo{author}{Cole, G.~D.}, \bibinfo{author}{Bai, Y.},
  \bibinfo{author}{Aspelmeyer, M.} \& \bibinfo{author}{Fitzgerald, E.~A.}
\newblock \bibinfo{title}{Free-standing Al(x)Ga(1-x)As heterostructures by
  gas-phase etching of germanium}.
\newblock \emph{\bibinfo{journal}{Appl. Phys. Lett.}}
  \textbf{\bibinfo{volume}{96}}, \bibinfo{pages}{261102}
  (\bibinfo{year}{2010}).

\bibitem{Cole10II}
\bibinfo{author}{Cole, G.~D.} \emph{et~al.}
\newblock \bibinfo{title}{Megahertz monocrystalline optomechanical resonators
  with minimal dissipation}.
\newblock In \emph{\bibinfo{booktitle}{Proc. IEEE Micro Electro Mechanical
  Syst.}}, \bibinfo{pages}{847--850} (\bibinfo{year}{2010}).
\newblock \bibinfo{note}{23rd Int. Conf.}

\end{thebibliography}

\end{document}